\documentclass[A4,12pt]{article}
\usepackage{epsfig}
\newcommand{\z}{&&\hspace*{-1cm}}

\newcommand{\bea}{\begin{eqnarray}}
\newcommand{\eea}{\end{eqnarray}}
\newcommand{\be}{\begin{equation}}
\newcommand{\ee}{\end{equation}}

\title{
Longitudinal structure function $F_L$
at small $x$ extracted
from the Berger-Block-Tan parametrization of $F_2$
  \author{ L.P.~Kaptari$^{1,2}$, A.V.~Kotikov$^{1,2}$, N.Yu.~Chernikova$^{3}$,
  Pengming Zhang$^{1,4}$ \\
    $^{1}$ Institute of Modern Physics, Lanzhou 730000, China\\
 $^{2}$ Bogoliubov Lab. Theor.  Phys.,  JINR, Dubna 141980, Russia\\
 $^{3}$ Sunday school, 141980, Dubna, Russia \\
 $^{4}$  
 University of Chinese Academy of Sciences, Beijing 100049, China
 }}

\begin{document}
\maketitle

\begin{abstract}
The longitudinal structure function $F_L(x,Q^2)$ is extracted
at low values of the Bjorken variable $x$ from
the Berger-Block-Tan parametrization of $F_2(x,Q^2)$.
The obtained $F_L(x,Q^2)$  does not violate the  high-energy asymptotic Froissart boundary
and is in a reasonable good agreement with the available experimental data.

\end{abstract}


\section{Introduction}

The nonperturbative corrections in the deep inelastic structure functions (SF) at
small values of the Bjorken  variable $x$  were expected to play an important role.
 However, it has been observed that even in the region of low momentum transfer $Q^2 \sim 1$ GeV$^2$,
 where traditionally the soft processes were considered to govern the cross sections,
 the perturbative QCD (pQCD) methods have been found to be quite adequate in description of
 such processes at moderate and low $x$, cf. Ref.~\cite{CooperSarkar:1997jk}.
 It should be noted, nonetheless,  that at extremely low $x$, $x\to 0$, the pQCD evolution provides
 a rather singular behaviour of the parton distribution functions (PDF) (see e.g. Ref.~\cite{Kotikov:1998qt} and
 references therein quoted),  which strongly violates the Froissart boundary~\cite{Froissart:1961ux}.
In Ref.~\cite{Berger:2007vf}  E.L. Berger, M.M. Block and C.I. Tan  have suggested a new parametrization
(in what follows referred to as the BBT parametrization)
of the SF $F_2(x,Q^2)$ (see also Refs.~\cite{Block:2011xb,Block:2014kza})
which describes fairly well the available experimental data on the reduced cross sections
and, at asymptotically low $x$, provides a behaviour of the cross sections  $\sim \ln^2 1/x$,
in an agreement with the Froissart predictions~\cite{Froissart:1961ux}.
This parametrization is also relevant in investigations of ultra-high energy processes,
such as scattering of cosmic neutrino off hadrons
(see \cite{Block:2011xb,Block:2014kza,Illarionov:2011wc}). It should be noted that, in case of neutrino
scattering other two SF's, the pure valence distribution $F_3(x,Q^2)$ and the longitudinal
SF $F_L(x,Q^2)$, are relevant to describe the process. While at low values of $x$ the valence
distribution $F_3(x,Q^2)$ vanishes, the longitudinal SF $F_L(x,Q^2)$ becomes predominant.
Thus, a theoretical analysis of the longitudinal SF $F_L(x,Q^2)$ at low $x$,
in context of fulfilment of the  Froissart booundary,
is  of a great importance also in ultra-high energy processes.

  In the present paper we investigate the behavior of
  the longitudinal SF $F_L(x,Q^2)$ at small $x$ by employing  the parametrization of
   $F_2(x,Q^2)$  presented in Ref.~\cite{Block:2014kza} within
   an approach previously suggested in Refs.~\cite{Kotikov:1993xe,Kotikov:1994vb,Kotikov:1994jh,Kotikov:1996yp}.
We demonstrate that, the small $x$ behavior of   $F_L(x,Q^2)$
can be directly  related to
 $F_2(x,Q^2)$ and, consequently, can be determined via   known parametrization~\cite{Block:2014kza}.

\section{Approach}

At the leading order (LO) of perturbation theory
the SF  $F_2$ is expressed through
the quark density $xf_q(x,Q^2)$ as
\footnote{The nonsinget quark density leads a negligible contribution at low $x$ values and will be neglected in
the present analysis.}
\be
F_2(x,Q^2) = e xf_q(x,Q^2),~~~ e= \sum^{n_f}_{i=1} e_i^2/n_f,~~~
\label{F2}
\ee
where $e$ is the average charge square
and $n_f$ is the number of flavors.

The longitudinal  $F_L(x,Q^2)$ relates   both the  quark and gluon densities, which obey in-turn
the famous DGLAP equations \cite{Gribov:1972ri}
\bea
\z F_L(x,Q^2)  = a_s(Q^2) e \sum_{a=s,g}
B^{(0)}_{L,a} (x) \otimes xf_a(x,Q^2), ~~f_1(x) \otimes f_2(x) \equiv \int^1_x \frac{dy}{y}
 f_1(y) f_2\left(\frac{x}{y}\right) , \label{FL} \\
\z \frac{d (xf_a(x,Q^2))}{dlnQ^2}  = -\frac{a_s(Q^2)}{2} \sum_{a,b=s,g} P^{(0)}_{ab}
(x) \otimes  xf_b(x,Q^2),~~ a_s(Q^2)
=  \frac{1}{\beta_0\ln(Q^2/\Lambda^2_{\rm LO})},
\label{DGLAP}
\eea
where $q^2=-Q^2$ and $x=Q^2/2pq$ ($p$ being the momentum of the  nucleon) denote  the  momentum transfer and the Bjorken scaling variable, respectively, $a_s(Q^2)=\alpha_s(Q^2)/(4\pi)$ is the QCD  running coupling,
$P^{(0)}_{ab} (x)$ $(a,b=q,g)$,
 and $B^{(0)}_{L,a} (x)$,  are the LO splitting  and  coefficient functions, respectively,
 and  $\beta_0$  is the first coefficient of the  QCD $\beta$-function.
 The symbol $\otimes$ stands for a shorthand notation of the convolution formula.

 Equations (\ref{F2}),  (\ref{FL}) and (\ref{DGLAP})
lead to the following  relations (in the leading order)
\bea
\z    \frac{dF_2(x,Q^2)}{dlnQ^2}  = -\frac{a_s(Q^2)}{2} \biggl[
e
P^{(0)}_{qg} (x) \otimes xf_g(x,Q^2) +
P^{(0)}_{qq} (x)  \otimes
F_2(x,Q^2)
 \biggr],
\label{dF2} \\
\z F_L(x,Q^2)  = a_s(Q^2) \biggl[ e B^{(0)}_{L,g} (x) \otimes xf_g(x,Q^2)
+ B^{(0)}_{L,q} (x) \otimes F_2(x,Q^2)
 \biggr] . \label{FL.1}
\eea

The  system  of equations (\ref{dF2})-(\ref{FL.1})  could be solved with respect to two unknown quantities,
say, $f_g(x,Q^2)$ and $F_L(x,Q^2)$, if   the third one (in our case the SF
$F_2(x,Q^2)$) were
known  from some independent considerations, e.g. from known experimental data.
In the present paper we suggest to use the parametrization of $F_2(x,Q^2)$ by Berger-Block-Tan~\cite{Block:2011xb,Block:2014kza} (in what follows
 referred to as
 $F_2^{BBT}(x,Q^2)$)
 which describes fairly well the recent experimental data~\cite{Aaron:2009aa}
and which, at $x\to 0$, obeys the Froissart requirements~\cite{Froissart:1961ux}.
In the region of our interests, i.e. for
$x <
0.01$ and  $0.15$ GeV$^2 < Q^2 < 3000$ GeV$^2$,  the proton $F_2^{BBT}(x,Q^2)$ reads as~\cite{Block:2014kza}
\begin{eqnarray}
F_{2}^{\rm BBT}(x,Q^2) = D(Q^2)
(1-x)^{\nu} \, \sum_{m=0}^2 A_m(Q^2) L^m \, ,
\label{n9}
\end{eqnarray}
where the logarithmic terms $L$ are defined  as
\bea
&&L= \ln \frac{1}{x} + L_1,~~ L_1= \ln \left(\frac{Q^2}{Q^2 + \mu^2}\right),\quad L_2= \ln \left(\frac{Q^2 + \mu^2}{\mu^2}\right)
\label{log}
\eea
and the remaining ones are as follows
\begin{eqnarray} &&
D(Q^2) = \frac{Q^2(Q^2 + \lambda M^2)}{(Q^2 + M^2)^2},\quad
A_i(Q^2) = \sum_{k=0}^2 a_{ik} \, L_2^k,~~
i=(0,1,2),~~ a_{02}=0,~~
\label{n9.0}
\end{eqnarray}
with
\be
\mu^2=2.82 \pm 0.29 \ {\mbox GeV}^2,~~ M^2=0.753 \pm 0.008 \ {\mbox GeV}^2,~~ \nu=11.49 \pm 0.99,~~
\lambda = 2.430 \pm 0.153,
\label{n9.11}
\ee
and
\bea
&&a_{00} = 0.255 \pm 0.016,~~
a_{01} \cdot 10^{1} = 1.475 \pm 0.3025,~~ \nonumber \\
&&a_{10} \cdot 10^{4} = 8.205 \pm 4.620,~~
a_{11} \cdot 10^{2} = -5.148 \pm 0.819,~~
a_{12} \cdot 10^{3} = -4.725 \pm 1.010, \nonumber \\
&&a_{20} \cdot 10^{3} = 2.217 \pm 0.142,~~
a_{21} \cdot 10^{2}=
1.244 \pm 0.086,~~
a_{22} \cdot 10^{4} =
5.958 \pm 2.320 \, .
\label{n10}
\eea

Now, by considering  $F_2(x,Q^2)=F_2^{BBT}(x,Q^2)$ as known, we are in a position to solve the
system (\ref{dF2})-(\ref{FL.1}) with respect to the  longitudinal SF
$F_L(x,Q^2)$. For the sake of consistency with the Froissart boundary condition, the SF
$F_L(x,Q^2)$ is sought in the same form as  $F_2^{BBT}(x,Q^2)$, cf. Eq.~(\ref{n9}),
\begin{eqnarray}
F_L^{\rm BBT}(x,Q^2) = (1-x)^{\nu} \, \sum_{m=0}^2 C_m(Q^2) L^m,\quad
C_i(Q^2) = \sum_{k=0}^2 c_{ik} L_2^k,\quad
i=(0,1,2).
\label{n9.1}
\end{eqnarray}
It should be noted that, in spite of  at the first glance  the above equations are relatively simple, direct solving of (\ref{dF2})-(\ref{FL.1}) turns to be rather complicate and cumbersome. However, one can substantially simplify the calculations if one considers Eqs.~(\ref{dF2})-(\ref{FL.1})
 in the space of Mellin momenta, by taking advantage of the fact the
convolution form $f_1(x)\otimes f_2(x)$ in  $x$ space becomes merely a product of individual Mellin momenta of the corresponding functions in the  space of Mellin momenta.

\subsection{Mellin transforms}
The Mellin transform of the ingredients in Eqs.~(\ref{dF2})-(\ref{FL.1}),
    $F_k(x,Q^2)$ (hereafter $k= 2,L $), $f_g(x,Q^2)$, $P^{(0)}_{ab} (x)$
    and $B^{(0)}_{La} (x)$  are defined as
\bea
&&M_k(n,Q^2) = \int^1_0 dx x^{n-2} F_k(x,Q^2),~~~
M_g(n,Q^2) = \int^1_0 dx x^{n-1} f_g(x,Q^2)\, ,
\label{Mg}\\
&&\gamma^{(0)}_{ab} (n) = \int^1_0 dx x^{n-2}
P^{(0)}_{ab} (x),~~~ B^{(0)}_{La} (n) = \int^1_0 dx x^{n-2}
B^{(0)}_{La} (x),~~ (a,b=q,g) \, ,
\label{gamma}
\end{eqnarray}
 which result in the following system of  equations for the corresponding momenta
\bea
\z \frac{dM_2(n,Q^2)}{dlnQ^2}  = -\frac{a_s(Q^2)}{2} \biggl[
e \gamma^{(0)}_{qg} (n) M_g(n,Q^2) + \gamma^{(0)}_{qq} (n)
M_2(x,Q^2)
 \biggr] ,
\label{dM2}  \\
\z M_L(n,Q^2)  = a_s(Q^2) \biggl[e B^{(0)}_{L,g} (n) M_g(x,Q^2)
+ B^{(0)}_{L,q} (n) M_2(x,Q^2)
 \biggr] , \label{FL.2}
\eea
where the  LO anomalous dimensions $\gamma^{(0)}_{qa} (n)$
and the Wilson  coefficients $ B^{(0)}_{L,a} (n)$
($a=q,g$) are

 \bea
\z \gamma^{(0)}_{qg} (n) = - \frac{4n_f (n^2+n+2)}{n(n+1)(n+2)},~~~
\gamma^{(0)}_{qq} (n) ~=~ \frac{32}{3} \biggl[\Psi(n)-\Psi(1)
-\frac{3}{4} + \frac{1}{2n} + \frac{1}{2(n+1)}\biggr],
\label{ana} \\
\z B^{(0)}_{L,g} (n) =  \frac{8n_f}{(n+1)(n+2)},~~~
B^{(0)}_{L,q} (n) ~=~  \frac{16}{3(n+1)}
\label{BL}
\eea
with $\Psi(n)$ as the Euler $\Psi$-function.

Eventually, inserting Eqs.~(\ref{ana}) and (\ref{BL}) in to  (\ref{dM2}) and (\ref{FL.2})
one obtains (cf., also Ref.~\cite{Kotikov:1994jh})
\be
M_L(n,Q^2)  =  b^{(0)}_{L,d} (n) \frac{dM_2(n,Q^2)}{dlnQ^2}
+ a_s(Q^2) \, b^{(0)}_{L,q} (n) M_2(x,Q^2),
\label{FL.2n}
\ee
where
\be
b^{(0)}_{L,d} (n) = - \frac{2 B^{(0)}_{L,g} (n)}{\gamma^{(0)}_{qg} (n)},~~
b^{(0)}_{L,q} (n) = B^{(0)}_{L,q} (n) - B^{(0)}_{L,g} (n)\, \frac{\gamma^{(0)}_{qq} (n)}{\gamma^{(0)}_{qg} (n)}
\label{bL}
\ee

\subsection{The  BBT Mellin transforms }
Expression  (\ref{FL.2n}) is the desired equation to be solved to find the parameters $C_i(x,Q^2)$ in Eq.~(\ref{n9.1}) and, consequently, to reconstruct the longitudinal
SF $F_L(x,Q^2)$.
Inserting (\ref{n9}) and (\ref{n9.1}) in to (\ref{Mg}) and   restricting oneself to small
values of $x$, i.e.   considering only the first Mellin momenta
$n=1+\omega$ with  $\omega \to 0$ (the limit $\omega \to 0$ corresponds to low
  values of $x$), one obtains  the desired Mellin transforms  $M_k(n,Q^2)$. Explicitly they read as
\bea
M_{2}^{\rm BBT}(n,Q^2) &=& D(Q^2) \, \sum_{m=0}^2 A_m(Q^2) P_m(\omega,\nu,L_1) + O(\omega)
\, , \label{M2BBT} \\
M_{L}^{\rm BBT}(n,Q^2) &=&  \sum_{m=0}^2 C_m(Q^2) P_m(\omega,\nu,L_1) + O(\omega) \, , \label{MLBBT}
\eea
where
\bea
&&P_0(\omega,\nu)=\frac{1}{\omega} - Z_1(\nu) \, ,~~
P_1(\omega,\nu,L_1)= \frac{1}{\omega^2} - Z_2(\nu) + L_1 P_0(\omega,\nu) \, ,  \nonumber \\
&&P_2(\omega,\nu,L_1)= 2 \left(\frac{1}{\omega^3} - Z_3(\nu)\right)  + 2L_1 P_1(\omega,\nu,L_1=0) +  L_1^2 P_0(\omega,\nu) \,
\label{Pi}
\eea
and
\be
Z_1(\nu)=S_1(\nu),~~Z_2(\nu)=\frac{1}{2} S_1^2(\nu) - \frac{1}{2} S_2(\nu), ~~
Z_3(\nu)=\frac{1}{6} S_1^2(\nu) - \frac{1}{2} S_1(\nu) S_2(\nu) + S_3(\nu) \, .
\label{Zi}
\ee

In the above equations the quantities $S_i(\nu)$ are related with   the Euler $\Psi(1+\nu)$-function and its derivatives $\Psi^{(m)}(\nu)=d^m/(d\nu)^m\Psi(\nu)$ as
\be
S_1(\nu)= \Psi(1+\nu) +\gamma_{\rm E},~~S_2(\nu)= \zeta_2 - \Psi^{(1)}(1+\nu),~~
S_2(\nu)=\frac{1}{2} \, \Bigl(\Psi^{(2)}(1+\nu) - \zeta_3 \Bigr) \, ,
\label{Si}
\ee
where $\gamma_{\rm E}$ is the Euler constant and $\zeta_i$ are the  Euler-Riemann  $\zeta$-functions.

\section{Determining the longitudinal
$\bf F_L^{\rm \bf BBT}(x,Q^2)$}
The further strategy of finding the quantities $C_i(Q^2)$ in  Eq.~(\ref{MLBBT}) is as following:
inserting Eqs.~(\ref{Pi})-(\ref{Si}) in to (\ref{FL.2n}) and equating the coefficients in front of
$1/\omega$-singularities in both sides of Eq. (\ref{FL.2n}) one finds the sought coefficients $C_i(Q^2)$
expressed via the BBT parameters (\ref{n9.11})-(\ref{n10}) and
momentum transfer $Q^2$.
After some algebra, the final result is~\footnote{In a full analogy with
the extraction \cite{Chernikova:2016xwx} of the gluon density.}
\bea
C_{2} &=& \hat{A}_2 +  \frac{8}{3} a_s D A_2 \, ,~~
C_{1} = \hat{A}_1 +  \frac{1}{2} \, \hat{A}_2 +  2D \left[ \frac{\mu^2}{Q^2+\mu^2} A_2 +
  \frac{4}{3} a_s \Bigl(A_1 + \left(4\zeta_2 - \frac{7}{2} \right) A_2 \Bigr) \right] \, ,
  \nonumber \\
C_{0} &=& \hat{A}_0 +  \frac{1}{4} \, \hat{A}_2 -\frac{7}{8} \, \hat{A}_2 +
D \Biggl[ \frac{\mu^2}{Q^2+\mu^2} \left(A_1 +\frac{1}{2} \, A_2 \right) \nonumber \\
 && +
  \frac{8}{3} a_s \left(A_0 + \left(2\zeta_2 - \frac{7}{4} \right) A_1 + \left(\zeta_2 - 4\zeta_3 +\frac{17}{8} \right)
    A_2 \right)
  \Biggr] \, ,
\label{C0}
\eea
where
\be
\hat{A}_i = \tilde{D} A_{i} + D \overline{A}_i \, \frac{Q^2}{Q^2+\mu^2},~~
\overline{A}_i(Q^2) = a_{i1} + 2 a_{i2} \, L_2,~~
i=(1,2,3),~~ a_{02}=0.
\label{hAiM}
\ee

\begin{figure}[t]
\centering
\vskip 0.5cm
\includegraphics[height=0.4\textheight,width=0.7\hsize]{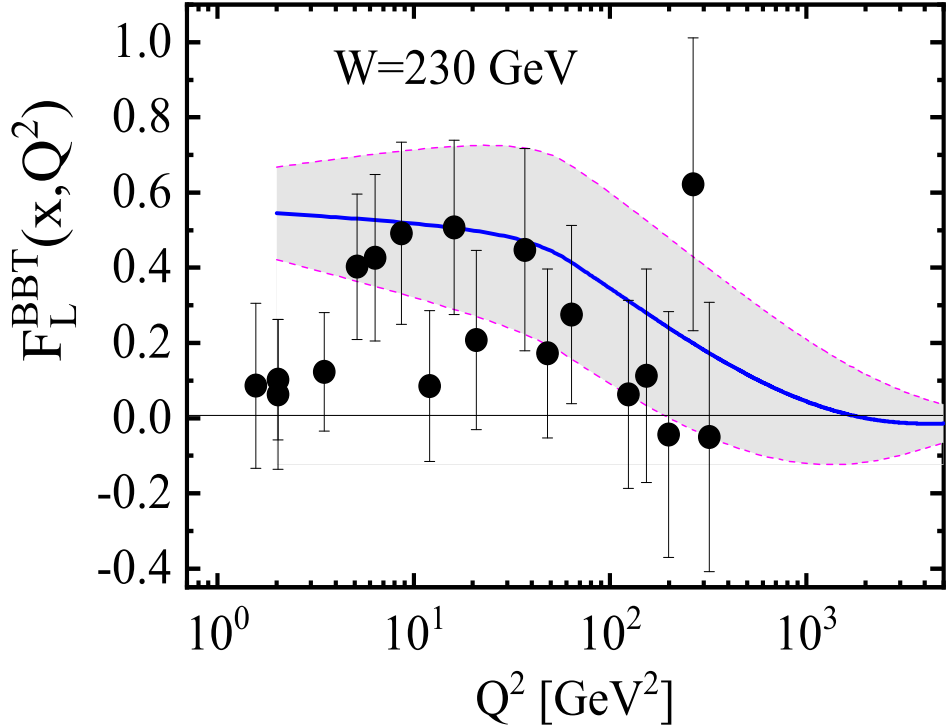}
\vskip -0.3cm
\caption{$Q^2$ dependence of the extracted longitudinal SF
$F_L^{\rm BBT}(x,Q^2)$
at fixed value of the invariant mass
$W$=230 GeV, solid curve.
The shaded area  corresponds to uncertainties of the BBT parameters, cf. Eq. (\ref{n10}).
Experimental data by the  H1 Collaboration is  taken from Ref.~\cite{Andreev:2013vha}.
}\label{fig:F2}
\end{figure}

In Fig.~\ref{fig:F2} we present  an example of the extracted SF
$F_L^{BBT}(x,Q^2)$
at a fixed value of the invariant mass $W^2=(p+q)^2$
in comparison with the available experimental data of the H1-collaboration taken from Ref.~\cite{Andreev:2013vha}.
The shaded area represents the uncertainties in the BBT parameters, cf. Eq. (\ref{n10}). It can be seen
that the suggested extraction procedure describes fairly well the data, especially in the interval $Q^2 >  5\ {\rm GeV}^2$.
At  lower values of the momentum transfer $Q^2 < 5\ {\rm GeV}^2$ the next-to-leading (NLO) corrections and their resummation
(see, e.g. Ref.~\cite{Kotikov:1993yw}) become rather important.
An analogous analysis of the $F_L^{BBT}(x,Q^2)$ at low $x$ and  $Q^2$ with NLO taken into account will be presented elsewhere.

\section{Summary}

In summary, we
determine the longitudinal SF
$F_L^{\rm BBT}(x,Q^2)$ from the Berger-Block-Tan parametrization of   $F_{2}^{\rm BBT}(x,Q^2)$ at low values of the Bjorken variable $x$.
The corresponding structure functions obey exactly the Froissart boundary restrictions
and can be used in investigations of ultra-high-energy scattering. Our calculations have been performed in the LO of pQCD.
As a next step, we plan to take into account  the NLO
corrections in a similar manner as in Ref.~\cite{Kotikov:1996yp}.  Since such corrections
are predicted to be large and negative,  their resummation~\cite{Kotikov:1993yw}
  is expected to improve substantially the agreement of the extracted $F_L^{\rm BBT}(x,Q^2)$
  with experimental data in the region of small $Q^2$,   $Q^2 \sim$ 1\ GeV$^2$.\\


Support by the National Natural Science Foundation of China (Grants
No. 11575254) is gratefully acknowledged. LPK and AVK highly appreciate the warm hospitality
at the Institute of Modern Physics where a bulk of the work has been done
and thank the CAS President's International Fellowship Initiative
(Grant  No.~2018VMA0029 and
No.~2017VMA0040) for support.
The work of AVK was in part supported by the RFBR Foundation through the Grant No.~16-02-00790-a.


\begin{thebibliography}{99}

\bibitem{CooperSarkar:1997jk}
  A.~M.~Cooper-Sarkar, R.~C.~E.~Devenish and A.~De Roeck,
  Int.\ J.\ Mod.\ Phys.\ A {\bf 13}, 3385 (1998);
  A.~V.~Kotikov,
  Phys.\ Part.\ Nucl.\  {\bf 38}, 1 (2007)
  [Phys.\ Part.\ Nucl.\  {\bf 38}, 828 (2007)].

\bibitem{Kotikov:1998qt}
  A.~V.~Kotikov and G.~Parente,
  Nucl.\ Phys.\  B {\bf 549}, 242 (1999);
  A.~Y.~Illarionov, A.~V.~Kotikov and G.~Parente Bermudez,
  Phys.\ Part.\ Nucl.\  {\bf 39}, 307 (2008).



\bibitem{Froissart:1961ux}
  M.~Froissart,
  Phys.\ Rev.\  {\bf 123}, 1053 (1961).




\bibitem{Berger:2007vf}
  E.~L.~Berger, M.~M.~Block and C.~I.~Tan,
  Phys.\ Rev.\ Lett.\  {\bf 98}, 242001 (2007);
  M.~M.~Block, E.~L.~Berger and C.~I.~Tan,
  Phys.\ Rev.\ Lett.\  {\bf 97}, 252003 (2006).


\bibitem{Block:2011xb}
  M.~M.~Block, L.~Durand, P.~Ha and D.~W.~McKay,
  Phys.\ Rev.\ D {\bf 84}, 094010 (2011).


\bibitem{Block:2014kza}
  M.~M.~Block, L.~Durand and P.~Ha,
  Phys.\ Rev.\ D {\bf 89}, no. 9, 094027 (2014)

\bibitem{Illarionov:2011wc}
  A.~Y.~Illarionov, B.~A.~Kniehl and A.~V.~Kotikov,
  Phys.\ Rev.\ Lett.\  {\bf 106}, 231802 (2011).


\bibitem{Kotikov:1993xe}
  A.~V.~Kotikov,
  Phys.\ Rev.\  D {\bf 49}, 5746 (1994);
  Phys.\ Atom.\ Nucl.\  {\bf 57}, 133 (1994.

\bibitem{Kotikov:1994vb}
  A.~V.~Kotikov,
  JETP Lett.\  {\bf 59}, 667 (1994);
  A.~V.~Kotikov and G.~Parente,
  Phys.\ Lett.\ B {\bf 379}, 195 (1996).

\bibitem{Kotikov:1994jh}
  A.~V.~Kotikov,
  J.\ Exp.\ Theor.\ Phys.\  {\bf 80}, 979 (1995);
  A.~V.~Kotikov and G.~Parente,
  J.\ Exp.\ Theor.\ Phys.\  {\bf 85}, 17 (1997)

\bibitem{Kotikov:1996yp}
  A.~V.~Kotikov and G.~Parente,
  Mod.\ Phys.\ Lett.\ A {\bf 12}, 963 (1997);
  G.~R.~Boroun,
  Phys.\ Rev.\ C {\bf 97}, no. 1, 015206 (2018).



\bibitem{Gribov:1972ri}
  V.~N.~Gribov and L.~N.~Lipatov,
  Sov.\ J.\ Nucl.\ Phys.\  {\bf 15}, 438 (1972)
  [Yad.\ Fiz.\  {\bf 15}, 781 (1972)];
  L.~N.~Lipatov,
  Sov.\ J.\ Nucl.\ Phys.\  {\bf 20}, 94 (1975)
  [Yad.\ Fiz.\  {\bf 20}, 181 (1974)];
  G.~Altarelli and G.~Parisi,
  Nucl.\ Phys.\ B {\bf 126}, 298 (1977);
  Y.~L.~Dokshitzer,
  Sov.\ Phys.\ JETP {\bf 46}, 641 (1977)
  [Zh.\ Eksp.\ Teor.\ Fiz.\  {\bf 73}, 1216 (1977)].

\bibitem{Aaron:2009aa}
  F.~D.~Aaron {\it et al.} [H1 and ZEUS Collaborations],
  JHEP {\bf 1001}, 109 (2010).




\bibitem{Chernikova:2016xwx}
  N.~Y.~Chernikova and A.~V.~Kotikov,
  JETP Lett.\  {\bf 105}, 223 (2017);
  A.~V.~Kotikov,
  Phys.\ Atom.\ Nucl.\  {\bf 80}, no. 3, 572 (2017).



\bibitem{Andreev:2013vha}
  V.~Andreev {\it et al.} [H1 Collaboration],
  Eur.\ Phys.\ J.\ C {\bf 74}, no. 4, 2814 (2014).



\bibitem{Kotikov:1993yw}
  A.~V.~Kotikov,
  Phys.\ Lett.\ B {\bf 338}, 349 (1994)
  [JETP Lett.\  {\bf 59}, 1 (1995)].

\end{thebibliography}
\end{document}